\input{aipcheck}

\documentclass[
    ,final            
  ]
  {aipproc}

\layoutstyle{8x11single}
\newcommand{\vecp}[1]{\stackrel{\longrightarrow}{#1}}

\begin{document}

\title{Euler-Poisson-Newton approach in Cosmology}


\author{R. Triay}{
  address={Centre de Physique Th\'eorique
\footnote{Unit\'e Mixte de Recherche (UMR 6207) du CNRS, et des universit\'es Aix-Marseille I,
Aix-Marseille II et du Sud Toulon-Var. Laboratoire affili\'e \`a la FRUMAM (FR 2291).}\\
CNRS Luminy Case 907, 13288 Marseille Cedex 9, France},
email={triay@cpt.univ-mrs.fr}
}

\author{H. H. Fliche}{
  address={LMMT\footnote{UPRES~EA 2596}, Fac. des Sciences et Techniques de St J\'er\^ome\\
av. Normandie-Niemen, 13397 Marseille Cedex 20, France},
email={ hhfliche@free.fr}
}

\begin{abstract}
This lecture provides us with Newtonian approaches for the interpretation of two puzzling cosmological observations that are still discussed subject~: a bulk flow and a foam like structure in the distribution of galaxies. For the first one, we model the motions describing all planar distortions from Hubble flow, in addition of two classes of planar-axial distortions with or without rotation, when spatial distribution of gravitational sources is homogenous. This provides us with an alternative to models which assume the presence of gravitational structures similar to Great Attractor as origin of a bulk flow. For the second one, the model accounts for an isotropic universe constituted by a spherical void surrounded by a uniform distribution of dust. It does not correspond to the usual embedding of a void solution into a cosmological background solution, but to a global solution of fluid mechanics. The general behavior of the void expansion shows a huge initial burst, which freezes asymptotically up to match Hubble expansion. While the corrective factor to Hubble law on the shell depends weakly on cosmological constant at early stages, it enables us to disentangle significantly cosmological models around redshift $z\sim 1.7$. The magnification of spherical voids increases with the density parameter and with the cosmological constant. An interesting feature is that for spatially closed Friedmann models, the empty regions are swept out, what provides us with a stability criterion.

\end{abstract}

\maketitle


\section{Introduction}
On one hand, it is a matter of fact that the use of Newtons theory of gravity for understanding cosmology structures is easier than general relativity. On the other hand, it is however clear that such an approach must be taken with caution at large scales although it ``is much closer to general relativity (GR) than commonly appreciated''  \cite{RueedeStraumann97}. With this in mind, we focus on Newtonian interpretation of two cosmological structures that have been mentioned in the literature, namely~: the presence of a bulk flow and a foam like structure in the observed distribution of galaxies. For that, we write  Euler-Poisson equations system in adapted coordinates, what enables us to define in a straightforward way solutions, as Newton-Friedmann and the vacuum (de Sitter) models. The bulk flow is analysed within an an-isotropic extension of Hubble model, see (\cite{FST06}). Subsequently, the model of a single spherical void is derived within a covariant approach, see (\cite{FT06}). 

\section{Euler-Poisson equations system}\label{EPES}
As usual, for modelling the dynamics of the cosmological expansion, the distribution of gravitational sources is supposed to behave as dust ({\it i.e.}, such a description does not account for shear, collision, etc\ldots and the presence of radiations). The  motions of such a pressureless medium are described at position $\vec{r}$ and time $t$ by its specific density $\rho=\rho(\vec{r},t)$ and velocity $\vec{v}=\vec{v}(\vec{r},t)$. These fields are constraint by Euler equations system
\begin{eqnarray}
\frac{\partial\rho}{\partial\,t}+{\rm div}\left(\rho \vec{v}\right)&=&0\label{Euler2}\\
\frac{\partial \vec{v}}{\partial\,t}+\frac{\partial\vec{v}}{\partial\vec{r}}
\vec{v} &=& \vec{g} \label{Euler1}
\end{eqnarray}
where $ \vec{g}=\vec{g}(\vec{r},t)$ stands for the gravitational field. It satisfies the modified Poisson-Newton equations
\begin{eqnarray}
\vecp{\rm rot}\vec{g} &=&\vec{0}\label{Poisson1}\\
{\rm div}\vec{g} &=&-4\pi{\rm G}\rho+\Lambda \label{Poisson2}
\end{eqnarray}
where ${\rm G}$ is Newton constant of gravitation and $\Lambda$ the cosmological constant.

For the investigation of an expanding medium, it is convenient to write these equations with new coordinates, herein named {\it reference coordinates\/}. They are defined by 
\begin{equation}\label{tx}
(t,\quad \vec{x}=\frac{\vec{r}}{a}),\qquad a>0
\end{equation}
where $a=a(t)$ is a non constant monotonic function with $a_{\circ}=a(t_{\circ})=1$. Such a method is motivated by the proper dynamics of medium, so that the equations system which accounts for the motion translates into its simplest form.

From now, let ${\rm div}$ and $\vecp{\rm rot}$ denote the differential operators with respect to $\vec{x}$. Euler-Poisson-Newton (EPN) equations system reads in term of reference coordinates as follows
\begin{eqnarray}
\frac{\partial\rho_{\rm c}}{\partial\,t}+{\rm div}\left(\rho_{\rm c} \vec{v}_{\rm c}\right)&=&0\label{Euler2n}\\
\frac{\partial \vec{v}_{\rm c}}{\partial\,t}+\frac{\partial\vec{v}_{\rm c}}{\partial\vec{x}} \vec{v}_{\rm c} +2H\vec{v}_{\rm c} &=& \vec{g}_{\rm c} \label{Euler1n}\\
\vecp{\rm rot}\vec{g}_{\rm c} &=&\vec{0}\label{Poisson1n}\\
{\rm div}\vec{g}_{\rm c}  &=&-\frac{4\pi{\rm G}}{a^{3}}\rho_{\rm c}-3\left(\dot{H}+H^{2}-\frac{\Lambda}{3} \right)
\label{Poisson2n}
\end{eqnarray}
where the dotted variables stand for time derivatives,
\begin{eqnarray}\label{VetG}
\rho_{\rm c}=\rho a^{3},\qquad
\vec{v}_{\rm c}=\frac{{\rm d}\vec{x}}{{\rm d}t}=\frac{1}{a}\left(\vec{v}-H\vec{r}\right),\qquad H=\frac{\dot{a}}{a}
\end{eqnarray}
act respectively as the density and the velocity fields of medium in the {\em reference frame\/}, and
\begin{equation}\label{gC}
\vec{g}_{\rm c}=\frac{1}{a}\left(\vec{g}-\left( \dot{H}+H^{2}\right)\vec{r}\right)
\end{equation}
as the acceleration field. Let us note that this model depends on two dimensionless parameter
\begin{equation}\label{Parameters}
\Omega_{\circ}=\frac{8\pi G}{3H_{\circ}^{2}}\rho_{c},\qquad
\lambda_{\circ}=\frac{\Lambda}{3H_{\circ}^{2}}
\end{equation}
where $H_{\circ}=H(t_{\circ})$.

\subsection{Newton-Friedmann and Vacuum models}\label{solutions}
For these models, the most adapted family of functions $a=a(t)$ verify a Friedmann type equation
\begin{equation}\label{H}
\dot{H}+H^{2}-\frac{\Lambda}{3}+\frac{4\pi{\rm G}}{3}\rho_{\circ}a^{-3}=0
\end{equation}
where $\rho_{\circ}$ is an arbitrary constant which stands for a parameter of function $a(t)$. This differential equation admits the following integration constant
\begin{equation}\label{K}
K_{\circ}=\frac{8\pi{\rm G}}{3} \rho_{\circ}+\frac{\Lambda}{3}-H_{\circ}^{2}
\end{equation}
It integrates for providing us with the function $H(t)$ from
\begin{equation}\label{chrono}
H^{2}=\frac{\Lambda}{3}-\frac{K_{\circ}}{a^{2}}+\frac{8\pi{\rm G}}{3} \frac{\rho_{\circ}}{a^{3}}\geq 0
\end{equation}
Let us denote its asymptotical value
\begin{equation}\label{Hinfinity}
H_{\infty}=\lim_{a\to\infty}H=\sqrt{\frac{\Lambda}{3}}
\end{equation}
According to eq.\,(\ref{K}), $a$ depends on $H_{\circ}$ and on two additional parameters chosen among $\Lambda$, $\rho_{\circ}$ and $K_{\circ}$. At this step, excepted $\Lambda$, these variables do not identify to cosmological parameters associated to Friedmann solution, they intervene merely in the coordinates choice. The formal expression $t\mapsto a$ is derived as reciprocal mapping of a quadrature from eq.\,(\ref{chrono}), it stands for the reference frame chronology.

It is obvious that EPN equation system eq.~(\ref{Euler2n},\ref{Euler1n},\ref{Poisson1n},\ref{Poisson2n}) shows two trivial solutions~:
\begin{enumerate}
\item The one defined by 
\begin{equation}\label{FL}
\rho_{\rm c}=\rho_{\circ},\qquad
\vec{v}_{\rm c}=\vec{0}, \qquad
\vec{g}_{\rm c}=\vec{0}
\end{equation}
which accounts for a uniform distribution of dust, herein named Newton-Friedmann model (NF). According to eq.\,(\ref{chrono},\ref{VetG}), $a$ and $H$ recover now their usual interpretations in cosmology as  {\em expansion parameter\/} and {\em Hubble parameter\/} respectively, $\rho_{\circ}=\rho a^{3}$ identifies to the density of sources in the comoving space and $K_{\circ}$ interprets in GR as its scalar curvature. We limit our investigation to motions which do not correspond to cosmological bouncing solutions, what requires the constraint
\begin{equation}\label{contraint}
K_{\circ}^{3}<\left(4\pi{\rm G}\rho_{\circ}\right)^{2}\Lambda
\end{equation}
to be fulfilled, according to analysis on roots of third degree polynomials. Hence, the kinematics shows two distinct behaviors characterized by the sign of $K_{\circ}$. Indeed, $H$ decreases with time by reaching $H_{\infty}$ either upward ($K_{\circ}\leq 0$) or downward ($K_{\circ}>0$) from a minimum defined by
 \begin{equation}\label{extremum}
H_{m}=H_{\infty}\sqrt{1-\frac{K_{\circ}^{3}}{\Lambda\left(4\pi{\rm G} \rho_{\circ}\right)^{2}}}<H_{\infty}
\end{equation}
at (epoch) $a=4\pi{\rm G}\rho_{\circ}K_{\circ}^{-1}$, what defines a  {\it loitering period\/}.
\item  The one defined by
\begin{equation}\label{void}
\rho_{\rm c}=0,\qquad
\vec{v}_{\rm c}=\left( H_{\infty}-H \right) \vec{x}, \qquad
\vec{g}_{\rm c}=\frac{4\pi{\rm G}}{a^{3}}\rho_{\circ}\vec{x}
\end{equation}
which accounts for vacuum solution,  herein named Vacuum model (V).
\end{enumerate}

\section{Anisotropic Hubble model}\label{CVF}
This model accounts for {\em collisionless motions\/} which satisfy the following kinematics
\begin{equation}\label{HubbleLaw}
\vec{r}={\bf A}\vec{r}_{\circ},\qquad {\bf A}(t_{\circ})={\rm 1\mskip-4mu l}
\end{equation}
where ${\bf A}={\bf A}(t)$ stands for a non vanishing determinant matrix; its coefficients can be determined by an observer at rest with respect to Cosmological Background Radiation (CMB).  One has ${\rm det}{\bf A}>0$ because it identifies to a unit matrix at time $t=t_{\circ}$. From eq.\,(\ref{HubbleLaw}), the velocity field $\vec{v}=\vec{v}(\vec{r},t)$ reads
\begin{equation}\label{velocity}
\vec{v}=\frac{{\rm d}\vec{r}}{{\rm d}t}=\dot{{\bf A}}{\bf A}^{-1}\vec{r},\qquad 
\dot{{\bf A}}=\frac{{\rm d}}{{\rm d}t}{\bf A}
\end{equation}
where ${\bf A}^{-1}$ stands for the inverse matrix\footnote{${\bf A}^{-1}{\bf A}={\bf A}{\bf A}^{-1}={\rm 1\mskip-4mu l}$}. Let us choose
\begin{equation}\label{a}
a(t)=\sqrt[3]{{\rm det}{\bf A}}
\end{equation}
and with the unique matrix decomposition\footnote{By using the trace, the determinant one has $\frac{\rm
d}{{\rm d}t} {\rm det}{\bf A}={\rm Tr}\left(\dot{{\bf A}}{\bf A}^{-1}\right){\rm det}{\bf A}$.}
 \begin{equation}\label{H}
\dot{{\bf A}}{\bf A}^{-1}=H{\rm 1\mskip-4mu l}+\frac{H_{\circ}}{a^{2}}{\bf B},\qquad H=\frac{\dot{a}}{a},\qquad H_{\circ}=H(t_{\circ}),\qquad {\rm
tr}\;{\bf B}=0
\end{equation}
eq.(\ref{Poisson2n},\ref{VetG},\ref{gC}) provide us with the equations of motion in the reference frame
\begin{eqnarray}
\vec{v}_{\rm c}&=&\frac{H_{\circ}}{a^{2}}{\bf B}\vec{x}\label{vc1}\\
\vec{g}_{\rm c}&=& \frac{H_{\circ}}{a^{2}}\left(\dot{\bf B}+\frac{H_{\circ}}{a^{2}}{\bf B}^{2}\right)\vec{x} \label{gc1}\\
\frac{H_{\circ}^{2}}{a^{4}}{\rm tr}{\bf B}^{2}&=&-\frac{4\pi{\rm G}}{a^{3}}\rho_{\rm c}-3\left(\dot{H}+H^{2}-\frac{\Lambda}{3} \right) \label{dgc}
\end{eqnarray}
where $a$ stands for the (generalized) {\em expansion factor\/}, thus $H=H(t)$ acts as the usual Hubble factor,  and ${\bf B}={\bf B}(t)$ for a traceless matrix herein named {\em distortion matrix\/}. It
characterises the deviation from isotropy of the (dimensionless) velocity field, its amplitude is defined by the matrix norm
\begin{equation}\label{amplitude}
\|{\bf B}\|=\sqrt{{\rm tr}\left({\bf B}^{t}{\bf B}\right)}
\end{equation}
where the sign ''$^{\rm t}$'' stands for the matrix transposition. 
According to eq.\,(\ref{Parameters}), eq.\,(\ref{dgc}) reads as follows
\begin{equation}\label{Poisson2b}
\frac{1}{H_{\circ}^{2}}\left(\dot{H}+H^{2}\right)
=-\frac{\Omega_{\circ}}{2a^{3}}+\lambda_{\circ}-\frac{1}{3a^{4}}\beta_{2}
\end{equation}
where
 \begin{equation}\label{TrBn}
\beta_{n}(t)={\rm tr}\left({\bf B}^{n}\right),\qquad n=1,2,3
\end{equation}
Eq\,(\ref{Poisson2b}) shows that $\Omega_{\circ}$ does not depend on spatial coordinates, {\it i.e.\/} the space distribution of gravitational sources is necessarily {\em homogenous\/}
\begin{equation}\label{homogene}
\rho=\rho(t)
\end{equation}
and hence it stands for a constant parameter because of eq.\,(\ref{Euler2n}).

By multiplying each term of
eq.\,(\ref{Poisson2b}) by
$2\dot{a}a$, one identifies the following constant of motion
\begin{equation}\label{k}
\kappa_{\circ}=\frac{\Omega_{\circ}}{a}+\lambda_{\circ}a^{2}-\frac{\dot{a}^{2}}{H_{\circ}^{2}}-\frac{2}{3}\int^{a}_{1}\frac{\beta_{2}}{a^{3}}{\rm d}a = \Omega_{\circ}+\lambda_{\circ}-1
\end{equation}
It is important to note that the chronology\footnote{According to observations, one assumes $H_{\circ}>0$ since the case $H_{\circ}<0$ accounts for a collapse, what is not envisaged.}, which is given by
\begin{equation}\label{dt}
{\rm d}t=\frac{1}{H_{\circ}}\frac{a{\rm d}a}{\sqrt{P(a)}}
\end{equation}
where
\begin{equation}\label{Pr}
P(a)=\lambda_{\circ}a^{4}-\kappa_{\circ}a^{2}+\Omega_{\circ}a-\frac{2}{3}a^{2}\int^{a}_{1}\frac{\beta_{2}}{a^{3}}{\rm d}a\geq
0,\qquad P(1)=1
\end{equation}
is distortion dependent.

\subsection{A Class of possible motions}
Accordingly to usual Hubble model, if we assume a {\em radial\/} acceleration field
\begin{equation}\label{RadialAcceleration}
\vec{g}_{\rm c}\propto\vec{x}
\end{equation}
then the distortion matrix satisfies the evolution equation
\begin{equation}\label{gRadial}
\dot{\bf B}+\frac{H_{\circ}}{a^{2}}{\bf B}^{2}= \frac{1}{3}\frac{H_{\circ}}{a^{2}}{\rm tr}{\bf B}^{2}
\end{equation}
according to eq.\,(\ref{gc1}, \ref{dgc}). With
 \begin{equation}
{\rm d}\tau=H_{\circ}\frac{{\rm d}t}{a^{2}}\label{u}
\end{equation}
eq.\,(\ref{gRadial}) reads in a dimensionless form
\begin{equation}\label{Bstar00}
\frac{{\rm d}{\bf B}}{{\rm d}\tau}= \frac{1}{3}\beta_{2}{\rm 1\mskip-4mu l}-{\bf B}^{2}
\end{equation}
With eq.\,(\ref{vc1},\ref{gc1},\ref{dt},\ref{gRadial}), the equations of motion read
 \begin{eqnarray}
\frac{{\rm d}\vec{x}}{{\rm d}\tau}&=&{\bf B}\vec{x},\qquad
\frac{{\rm d}^{2}\vec{x}}{{\rm d}\tau^{2}}=\frac{1}{3}\beta_{2}\vec{x}\label{qvelocity}\\
{\rm d}\tau&=&\frac{{\rm d}a}{a\sqrt{P(a)}}\label{du}\label{t}
\end{eqnarray}
The resolution of these equations can be performed by mean of numerical techniques~: the evolution with time of distortion matrix ${\bf B}$ is derived from eq.\,(\ref{Bstar00}), the particles trajectories $\tau\mapsto\vec{x}(\tau)$ are obtained by integrating eq.\,(\ref{qvelocity}) and the evolution of the generalized expansion factor $a$ from eq.\,(\ref{t}).

\subsubsection{Analysis of analytic solutions}

Analytic solutions of eq.\,(\ref{qvelocity},\ref{t},\ref{Bstar00}) can be obtained thanks to particular
properties of distortion matrix ${\bf B}$. The parameters $\lambda_{\circ}$, $\Omega_{\circ}$ and $H_{\circ}$ given in eq.\,(\ref{Parameters}) correspond to cosmological parameters of Friedmann-Lema\^{\i}tre (FL) solution. The constraint $\beta_{2}=0$ in eq.\,(\ref{dt},\ref{Pr}) provides us with
the FL chronology, where $\kappa_{\circ}$ in eq.\,(\ref{k}) represents the curvature parameter in the FL model ({\em i.e.\/} the dimensionless scalar
curvature
$\Omega_{k}$ of the comoving space, see~\cite{TriayEtAl96}), while the flatness of (simultaneous events) Newton space. It must be noted that the particle
position $\vec{x}$ as defined in eq.\,(\ref{tx}) does not identify to the usual FL comoving coordinate because the (generalized) expansion factor $a$
depends on the anisotropy unless $\beta_{2}=0$. 

\paragraph{Evolution of functions $\beta_{n=2,3}$}
Because ${\bf B}$ is a traceless matrix, its characteristic polynomial reads
\begin{equation}\label{Polynome}
Q(s)={\rm det}\left(s{\rm 1\mskip-4mu l}-{\bf B}\right)=s^{3}-\frac{1}{2}\beta_{2}s-\frac{1}{3}\beta_{3}
\end{equation}
according to Leverrier-Souriau's algorithm \cite{Souriau92}. With Cayley-Hamilton's theorem ({\it i.e.} $Q({\bf B})=0$) and eq.\,(\ref{Bstar00}) we obtain
the following differential equations system
\begin{eqnarray}
\frac{{\rm d}}{{\rm d}\tau}\beta_{2}&=& -2\beta_{3} \label{TrB2}\\
\frac{{\rm d}}{{\rm d}\tau}\beta_{3}&=& -\frac{1}{2}\beta_{2}^{2} \label{TrB3}
\end{eqnarray}
and we note that the discriminant of third order polynomial $Q$, it is proportional to
\begin{equation}\label{c}
\alpha=3\beta_{3}^{2}-\frac{1}{2}\beta_{2}^{3}
\end{equation}
is a \underline{constant of motion} ({\it i.e.\/}, ${\rm d}\alpha/{\rm d}\tau=0$). The integration of eq.\,(\ref{TrB2},\ref{TrB3}) gives
\begin{equation}\label{TraceEvol}
\tau=\tau_{\circ}+\epsilon\frac{\sqrt{6}}{2}\int_{\beta_{2}(\tau_{\circ})}^{\beta_{2}(\tau)}\frac{{\rm d}x}{\sqrt{2\alpha+x^{3}}},\qquad \epsilon=\pm 1
\end{equation}
Hence, $\beta_{2}$ is defined by a quadrature, and $\beta_{3}$ from eq.\,(\ref{TrB2}); in addition of the singular solution
\begin{equation}\label{SingularSol}
\beta_{2}=\beta_{3}=0,\qquad (i.e.,\quad {\bf B}^{3}=0)
\end{equation}
defined equivalently either by $\beta_{2}=0$ or $\beta_{3}=0$, according to eq.\,(\ref{TrB2},\ref{TrB3}).

The related dynamics depends on roots $\eta_{i=1,2,3}$ of characteristic polynomial $Q$ given in eq.\,(\ref{Polynome}), {\it i.e.\/} the eigenvalues of
distortion matrix ${\bf B}$. Their real values identify to dilatation rates at time $\tau$ toward the corresponding (time dependent) eigenvectors (not
necessarily orthogonal). Their sum is null ($\beta_{1}=0$) and their product ($\beta_{3}=3{\rm det}{\bf B}$) is either decreasing
with time or is null, according to eq.\,(\ref{TrB3}). The sign of $\alpha$ given in eq.\,(\ref{c}) is used to classify the solutions as follows~:
\begin{itemize}
\item if $\alpha=0$ then $Q$ has a real double root $\eta_{1}=\eta_{2}$ and a simple one $\eta_{3}$. The related instantaneous kinematic shows a
planar-axial symmetry (either a contraction within a plane with an expansion toward a transverse direction or vice versa), see
sec.~\ref{PlanarAxialKinematic}. If $\eta_{1}=\eta_{3}$ then both vanish and the related solution identifies to the singular one defined in eq.\,(\ref{SingularSol});
\item if $\alpha>0$ then $Q$ has a single real root $\eta_{1}$; 
\item if $\alpha<0$ then $Q$ has three distinct real roots $\eta_{i=1,2,3}$. Their order is conserved during the evolution (since a coincidence of
eigenvalues makes $\alpha=0$), the largest one must be positive while the smallest one must be negative (because $\beta_{1}=0$).
\end{itemize}

\paragraph{Planar kinematics}\label{PlanarKinematic}
The singular solution ${\bf B}^{3}=0$ shows a FL chronology and the distortion matrix
\begin{equation}\label{distortionFree}
{\bf B}=-{\bf B}_{\circ}^{2}\tau+{\bf B}_{\circ},\qquad {\bf B}_{\circ}^{3}={\bf 0}
\end{equation}
is solely defined by its initial value ${\bf B}_{\circ}$, according to eq.\,(\ref{Bstar00}). It is neither
symmetric nor asymmetric (otherwise it vanishes), see eq.\,(\ref{amplitude}). Hence, eq.\,(\ref{qvelocity}) transforms
\begin{equation}\label{distortionFreeQ}
\frac{{\rm d}\vec{x}}{{\rm d}\tau}=\left(-{\bf B}_{\circ}^{2}\tau+{\bf B}_{\circ}\right)\vec{x}
\end{equation}
which accounts for eternal motions
\begin{equation}\label{distortionFreeQtriv}
\vec{x}=\exp{\left(-{\bf B}_{\circ}^{2}\frac{\tau^{2}}{2}+{\bf B}_{\circ}\tau\right)}\vec{x}_{\circ}
=\left(1+{\bf B}_{\circ}\tau\right)\vec{x}_{\circ}
\end{equation}
The trajectory of a particle located at initial position $\vec{x}_{\circ}$ identifies to a straight line toward the direction ${\bf
B}_{\circ}\vec{x}_{\circ}$. The analysis of ${\bf B}_{\circ}$ range ({\it i.e.\/}, its image) provides us with characteristics of trajectories
flow. The nilpotent property of ${\bf B}_{\circ}$ shows that its kernel is not empty ${\rm Ker}({\bf B}_{\circ})\neq\emptyset$. Its dimension ${\rm
dim}\left({\rm Ker}({\bf B}_{\circ})\right)=m$ characterizes the kinematics, which is either planar ($m=1$) or directional ($m=2$), {\it i.e.\/} a
bulk flow. Conversely, if the kernel of distortion matrix ${\bf B}$ is not empty then $\beta_{3}=3\,{\rm det}\left({\bf B}\right)=0$, and thus
$\beta_{2}=0$, see to eq.\,(\ref{TrB2},\ref{TrB3}). Therefore, all planar kinematics can be described by such a model.

\paragraph{Planar-Axial kinematics}\label{PlanarAxialKinematic}
If $\beta_{2}\neq0$ then the chronology differentiates from FL one. Let us focus on the $\alpha=0$ with two distinct eigenvalues
$\eta_{1}\neq \eta_{3}$ class of solutions. With eq.\,(\ref{TraceEvol}), eq.\,(\ref{TrB2},\ref{TrB3}) integrate
\begin{equation}\label{TraceBn}
\beta_{n}=\frac{6}{\left(\tau-\tau_{\star}\right)^{n}},\quad (n=2,3), \qquad \tau_{\star}=\tau_{\circ}+\sqrt{6}\beta_{2}^{-1/2}(\tau_{\circ})
\end{equation}
which shows a singularity at date $\tau=\tau_{\star}>0$ that splits the motion in two regimes $\tau<\tau_{\star}$ and $\tau>\tau_{\star}$. The complete
investigation of this singularity problem demands to solve an integro-differential equation, see eq.\,(\ref{dt},\ref{Pr}). The roots of $Q$
read
\begin{equation}\label{C0ValP}
\eta_{1}=\frac{1}{\left(\tau_{\star}-\tau\right)},\qquad
\eta_{3}=-2\eta_{1}
\end{equation}
where $\eta_{1}$ stands for the double root. Among others, two class of solutions are defined by mean of a constant (time independent) matrix ${\bf P}$,
the projector associated to $\eta_{1}$, see~\cite{Souriau92},
\begin{equation}\label{P1}
{\bf P}^{2}={\bf P},\quad
{\rm tr}{\bf P}=2
\end{equation}
They describe distinct kinematics depending on whether matrix ${\bf B}$ is diagonalizable.

\begin{itemize}
\item {\it Irrotational motions\/}~:
If ${\bf B}$ is diagonalizable then
\begin{equation}\label{PancakeB}
{\bf B}=\eta_{1}\left(3{\bf P}-2{\rm 1\mskip-4mu l}\right)
\end{equation}
From eq.\,(\ref{qvelocity},\ref{PancakeB}), one has
\begin{equation}\label{Caust_q_vel}
\frac{{\rm d}\vec{x}}{{\rm d}\tau}=\eta_{1}\left(3{\bf P}-2{\rm 1\mskip-4mu l}\right)\vec{x}\\
\end{equation}
and the solution reads
\begin{equation}\label{irrotational}
\vec{x}= -\eta_{1}{\bf P}\vec{\xi}+\frac{1}{\eta_{1}^{2}}\left({\rm 1\mskip-4mu l}-{\bf P}\right)\vec{\xi}
\end{equation}
where $\vec{\xi}$ is constant. If the eigenvectors are orthogonal then the kinematics accounts for irrotational motions. 

\item {\it Rotational motions\/} ~:
If ${\bf B}$ is not diagonalizable then
\begin{eqnarray}
{\bf B}&=&\eta_{1}\left(3{\bf P}-2{\rm 1\mskip-4mu l}\right)+\frac{1}{\eta_{1}^{2}}{\bf N}\label{PancakeB1}\\
\frac{{\rm d}\vec{x}}{{\rm d}\tau}&=&\left(\eta_{1}\left(3{\bf P}-2{\rm 1\mskip-4mu l}\right)+\frac{1}{\eta_{1}^{2}}{\bf N}\right)\vec{x}\label{PancakeB2}
\end{eqnarray}
where ${\bf N}$ is a constant nilpotent matrix, which accounts for rotational motions on the eigenplane of ${\bf P}$. The
solution reads
\begin{equation}\label{rotational}
\vec{x}= -\eta_{1}{\bf P}\vec{\xi}+\frac{1}{\eta_{1}^{2}}\left({\rm 1\mskip-4mu l}-{\bf P}\right)\vec{\xi}+\frac{1}{\eta_{1}^{2}}{\bf N}\vec{\xi}
\end{equation}
where $\vec{\xi}$ is constant.
\end{itemize}

\subsubsection{Discussion}
As a result, this anisotropic generalization requires an homogeneous distributions of matter. Because of the presence of strong density inhomogeneities in the sky distribution of galaxies catalogs, one is forced to ask whether it describes correctly the dynamics of observed cosmic structures. In principle, such a remark should be also sensible for questioning Hubble law when, regardless the isotropy, it is a fact that perturbations are not so dominant otherwise it would never have been highlighted. Actually, homogeneity is implicitly assumed in the standard cosmology for the interpretation of CMB isotropy and the redshift of distant sources, which provides us with an expanding background\footnote{Namely the comoving space of FL world model onto which the gravitational instability theory is applied for understanding the formation of
cosmic structures.}. It is with such a schema in mind that this anisotropic Hubble law provides us with an hint on the behavior of the cosmic flow from decoupling era up to present date in order to answer whether the observed bulk flow is due exclusively to tidal forces.

\subsection{Cosmic flow of flat LSS (${\bf B}^{3}=0$)}
The ${\bf B}^{3}=0$ class of solutions has interesting properties with regard to the stability of large scale structures that show a flat spatial
distribution. To answer the question of whether observations define unambiguously the kinematics, the distortion matrix ${\bf B}$ is decomposed as follows
 \begin{equation}\label{Bdecomp}
{\bf B}={\bf S}+{\bf j}(\vec{\omega}),\qquad {\rm tr}{\bf S}=0
\end{equation}
where ${\bf S}$ and ${\bf j}(\vec{\omega})$ stand for its symmetric and its asymmetric\footnote{The operator ${\bf j}$
stands for the vector product, $\vec{u}\times\vec{\omega}={\bf j}(\vec{u})(\vec{\omega})$.} component, and
\begin{equation}\label{tourbillon}
\vec{\omega}=\frac{a^{2}}{H_{\circ}}\vec{\sigma},\qquad \vec{\sigma}=\frac{1}{2}\vecp{\rm rot}\vec{v}
\end{equation}
accounts for the motion rotation, $\vec{\sigma}$ being the swirl vector.  Hence, eq.\,(\ref{distortionFree})
gives
\begin{equation}\label{tourbillonV}
{\bf B}\vec{\omega}={\bf S}\vec{\omega}
\end{equation}
The evolution of the anisotropy with time is defined by
\begin{eqnarray}
{\bf S}&=&-\left({\bf S}_{\circ}^{2}+ {\bf j}(\vec{\omega_{\circ}}){\bf j}(\vec{\omega_{\circ}})\right)\tau+{\bf S}_{\circ}\label{CinPlane1}\\
{\bf j}(\vec{\omega})&=&-\left({\bf S}_{\circ}{\bf j}(\vec{\omega_{\circ}})+{\bf j}(\vec{\omega_{\circ}}){\bf S}_{\circ}\right)\tau+{\bf
j}(\vec{\omega_{\circ}})\label{CinPlane2}
\end{eqnarray}
which couples the symmetric and the antisymmetric parts of the distortion matrix. The swirl magnitude reads
\begin{equation}\label{tourbillonM}
\omega=\sqrt{\langle\vec{\omega},\vec{\omega}\rangle}=\sqrt{\frac{1}{2}{\rm tr}{\bf S}^{2}}
\end{equation}
according to eq.\,(\ref{tourbillon}), since $\beta_{2}=0$. Its orientation cannot be determined from the data because the above equations describe two
distinct kinematics corresponding to $\pm\vec{\omega_{\circ}}$ that cannot be disentangle. According to eq.\,(\ref{Bdecomp}), if (and only if) the
rotation $\omega=0$ then the distortion vanishes ${\bf S}={\bf 0}$ since ${\bf B}$ is either a symmetric or antisymmetric. In other words, a planar
distortion has necessarily to account for a rotation.

\subsubsection{Constant distortion}
Among above solutions which show planar kinematics, let us investigate the (simplest) one defined by ${\bf
B}^{2}_{\circ}={\bf 0}$. In such a case, $\vec{k}_{\circ}\propto\vec{\omega}$ and ${\bf S}\vec{\omega}=\vec{0}$.  According
to eq.\,(\ref{CinPlane1},\ref{CinPlane2},\ref{tourbillonM}), linear calculus shows that the distortion is constant
\begin{equation}\label{tourbillonF}
{\bf S}={\bf S}_{\circ},\qquad \omega=\omega_{\circ}
\end{equation}
Such a distortion in the Hubble flows produces a rotating planar velocities field with magnitude $\propto H_{\circ}a^{-2}$. In the present case, the
model parameters can be easily evaluated from data. The observed cosmic velocity fields are partially determined by their radial component
\begin{equation}\label{VitRad}
v_{r}=\langle\vec{v},\frac{\vec{r}}{r}\rangle=cz,\qquad \vec{r}=\vec{r}(m)=r\vec{u}=a\vec{x},\qquad r=ct
\end{equation}
where $m$, $z$, $\vec{u}$, $t$ stand respectively for the apparent magnitude, the redshift, the line of sight, the photon emission date of
the galaxy and $c$ the speed of the light. According to eq.\,(\ref{velocity},\ref{H},\ref{Bdecomp}), the radial velocity of a galaxy
located at position $\vec{r}$ is given by
\begin{equation}\label{Vr}
v_{r}=\left(H +\tilde{H}_{\vec{u}}\right)r,\qquad \tilde{H}_{\vec{u}}=\frac{H_{\circ}}{a^{2}}\vec{u}\cdot{\bf
S}\vec{u}
\end{equation}
Because ${\rm tr}{\bf S}=0$, it is clear that the sum of three radial velocities $v_{r}$ corresponding to galaxies located in the sky toward
orthogonal directions and at same distance $r$ provides us with the quantity $H$. Hence, simple algebra shows that the sample average of radial
velocities within a sphere a radius $r$ is equal to
\begin{equation}\label{Hsample}
\langle v_{r}\rangle=H\langle \vec{r}\rangle
\end{equation}
Therefore, for motions described by eq.\,(\ref{HubbleLaw}), the statistics given in eq.\,(\ref{Hsample}) provides us with a genuine interpretation
of Hubble parameter $H=H(t)$. Hence, according to eq.\,(\ref{H},\ref{dt},\ref{Pr}), one obtains the (generalized) expansion factor
\begin{equation}\label{at}
a(t)=\exp{\int_{t_{\circ}}^{t}H(t){\rm d}t}
\end{equation}
Hence, the cosmological parameters can be estimated by fitting the data to the function 
\begin{equation}\label{psi}
\psi(t)_{\lambda_{\circ},\kappa_{\circ},\Omega_{\circ}}=\sqrt{P(a)}=a^{2}H/H_{\circ},\quad \lambda_{\circ}+\kappa_{\circ}+\Omega_{\circ}=1
\end{equation}
The component of matrix ${\bf S}$ can be estimated by substituting $H$ in eq.\,(\ref{Vr}), and $\omega$ is obtained from eq.\,(\ref{tourbillonF}).

It is clear that the above model is derived in the Galilean reference frame, where the Euler-Poisson equations system can be applied. Hence, a non
vanishing velocity of the observer with respect to this frame an produces a bipolar harmonic signal in the $\tilde{H}_{\vec{u}}$ distribution
of data in the sky, which can be (identified and then) subtracted. 

\subsection{Discussion}
The dynamics of a homogenous medium and anisotropic moving under Newton gravity was already studied by describing the evolution of an
ellipsoid \cite{ZeldovichNovikov83}. The current approach enables us to identify characteristics of the dynamics of the deformation from isotropic Hubble
law in a more systematic way by mean of the distortion matrix.

At first glance, if the planar anisotropic of the space distribution of galaxies within the Local Super Cluster (LSC) is stable then the above
solution can be used for understanding its cosmic velocity fields, $\vec{k}_{\circ}$ being orthogonal to LSC plane. It is well known however that
the distribution of galaxies is not so homogenous as that, whereas this model describes motions of an homogenous distribution of gravitational
sources. On the other hand, such an approximation level is similar to the one which provides us with the observed Hubble law, that is included in this model
(${\bf S}=0$). 

\section{Spherical  voids in Newton-Friedman universe}\label{DynamicsVoid}
For modelling the dynamics of a spherical void in a uniform dust distribution, we use a covariant formulation of Euler-Poisson equations system (see Appendix). The model is obtained by sticking together the local solutions V and NF of Euler-Poisson equations system as defined previously in the Newton-Friedmann and Vacuum models, where the function $a$ stands for the Friedmann expansion parameter. Such a model accounts for the dynamics of their common border ({\it i.e.\/} boundaries conditions), which is a {\it material shell\/}. For convenience in writing, the symbols S, V and NF denotes both the medium and the related dynamical model. A qualitative analysis of solutions is performed and a general discussion is given subsequently

\subsection{Dynamical model}\label{ModelVoid}
We consider three distinct media~:  a material shell with null thickness (S), an empty inside (V) and outside a uniform dust distribution (NF). These media behave such that S makes the juncture of V with NF as given by eq.\,(\ref{FL},\ref{void}). The tension-stress on S is assumed to be negligible, what is characterized  by a (symmetric contravariant) mass-momentum tensor defined as follows
\begin{equation}\label{Ts}
T_{\rm S}^{00}=(\rho_{\rm S})_{c},\quad
T_{\rm S}^{0j}=(\rho_{\rm S})_{c}v_{c}^{j},\quad
T_{\rm S}^{jk}=(\rho_{\rm S})_{c}v_{c}^{j}v_{c}^{k}
\end{equation}
The background is described by the following mass-momentum tensor
\begin{equation}\label{To}
T_{\rm NF}^{00}=\rho_{c},\quad
T_{\rm NF}^{0j}=0,\quad
T_{\rm NF}^{jk}=0
\end{equation}
According to Appendix~\ref{Covariant}, since the eulerian function(al)
\begin{equation}\label{T}
{\cal T}(x\mapsto\gamma)=\int T_{\rm S}^{\mu\nu} \gamma_{\mu\nu}{\rm d}t{\rm d}S+
\int T_{\rm NF}^{\mu\nu} \gamma_{\mu\nu}{\rm d}t{\rm d}V
\end{equation}
vanishes when $\gamma$ reads in the form $\gamma_{\mu\nu}=\frac{1}{2}\left( \hat{\partial}_{\mu} \xi_{\nu}+ \hat{\partial}_{\nu} \xi_{\mu}\right)$, one has
\begin{eqnarray}
&&\int_{\rm S}
\left(\left(\partial_{0}\xi_{0}+g_{c}^{j}\xi_{j}\right)+\left(\partial_{j}\xi_{0}+\partial_{0}\xi_{j}-2H\xi_{j}\right)v_{c}^{j} 
+ v_{c}^{j}v_{c}^{k}\partial_{j}\xi_{k}\right)(\rho_{\rm S})_{c}\,{\rm d}t x^{2}{\rm d}\Omega \nonumber\\
&=&- \int_{\rm NF} \rho_{c}\partial_{0}\xi_{0}\,{\rm d}t  x^{2} {\rm d}x{\rm d}\Omega \label{EqFond}
\end{eqnarray}
where ${\rm d}\Omega$ stands for the solid angle element. The radial symmetry of solutions enables us to write the {\em reduced\/} peculiar velocity and acceleration of a test particle located on the shell as follows
\begin{equation}\label{radial}
\vec{v}_{c}=\alpha\vec{x},\qquad
\vec{g}_{c}=\beta\vec{x}
\end{equation}
where the functions $\alpha=\alpha(t)$ and $\beta=\beta(t)$ have to be determined. A by part integration of eq.\,( \ref{EqFond}) provides us with
\begin{eqnarray}\label{EulerS}
&&
\int_{t_{1}}^{t_{2}}
\left(\partial_{0}(\rho_{\rm S})_{c}+3(\rho_{\rm S})_{c}\alpha-\rho_{c} \alpha x\right)x^{2}\xi_{0} {\rm d}t\\
&=&
\int_{t_{1}}^{t_{2}}
\left(\partial_{0}\left((\rho_{\rm S})_{c}\alpha\right)  +4(\rho_{\rm S})_{c}\alpha^{2} + 2H(\rho_{\rm S})_{c}\alpha-
(\rho_{\rm S})_{c}\beta x^{-1}\right)x^{3}\tilde{\xi}
{\rm d}t\nonumber
\end{eqnarray}
where $x=\|\vec{x}\|$ stands for the radius of S and $\tilde{\xi}=\sqrt{\xi_{1}^{2}+\xi_{2}^{2}+\xi_{2}^{2}}$.  This equality must be fulfilled for all bounded time interval and compact support 1-form. Hence, we easily derive the conservation equations for the mass
\begin{equation}\label{masseC}
\partial_{0}(\rho_{\rm S})_{c}+\left(3(\rho_{\rm S})_{c}-\rho_{c} x\right) \alpha = 0
\end{equation}
and for the momentum
\begin{equation}\label{impulsionC}
\frac{{\rm d}\alpha}{{\rm d}t}+\left(1+\frac{\rho_{c}}{(\rho_{\rm S})_{c}}x\right)\alpha^{2}+2H \alpha+ \frac{\beta}{x}=0
\end{equation}
With eq.\,(\ref{tx}), the calculation of the gravitational force from the entire shell acting on a particular point\footnote{The modified newtonian gravitation field reads
$
\vec{g}=\left(
\frac{\Lambda}{3}-\frac{{\rm G}m}{r^{3}}\right)\vec{r}
$}
provides us with
\begin{equation}\label{beta}
\beta=\frac{4\pi{\rm G}}{a^{3}}\left(\frac{\rho_{c}}{3}- \frac{(\rho_{\rm S})_{c}}{2x} \right)
\end{equation}
About mass conservation, it is noticeable that
\begin{equation}\label{bord}
(\rho_{\rm S})_{c}=\frac{1}{3}\rho_{c}x
\end{equation}
is solution of eq.\,(\ref{masseC}), what ensures that the amount of matter which forms the shell comes from its interior. Hence, eq.\,(\ref{impulsionC}) transforms
\begin{equation}\label{impulsionC1}
\frac{{\rm d}\alpha}{{\rm d}t}+4\alpha^{2}+2H \alpha- \frac{2\pi{\rm G}}{3} \frac{\rho_{c}}{a^{3}}=0
\end{equation}
It is convenient to use the dimensionless variable
\begin{equation}\label{chi}
\chi=4\frac{\alpha}{H_{\circ}}a^{2}
\end{equation}
where the ratio $\alpha\,H_{\circ}^{-1}$ stands for the expansion rate of S in the reference frame. Hence, eq.\,(\ref{impulsionC1}) transforms into a Riccati equation
\begin{equation}\label{impulsionC2}
\frac{{\rm d} \chi}{{\rm d}a}= \left(\Omega_{\circ}-\frac{\chi^{2}}{a} \right)
\frac{1}{\sqrt{P(a)}}
\end{equation}
where
\begin{equation}\label{P}
P(a)= \lambda_{\circ}a^{4}-k_{\circ}a^{2}+\Omega_{\circ} a, \qquad k_{\circ} = \frac{K_{\circ}}{H_{\circ} ^{2}},\qquad P(1)=1
\end{equation}
where the dimensionless parameters\footnote{These notations are preferred to the usual $\Omega_{\Lambda}=\lambda_{\circ}$ and $\Omega_{K}=-k_{\circ}$ for avoiding ambiguities on the interpretation of cosmological parameters, see {\it e.g.\/}, \cite{FlicheSouriau79},\cite{FlicheEtal82}.} are defined in eq.~(\ref{Parameters}, \ref{K}).
According to eq.\,(\ref{VetG},\ref{chi}), the evolution of the radius of S is given by
\begin{equation}\label{Croissance}
x=x_{i}\exp\left(\int_{a_{i}}^{a}\frac{\chi{\rm d}a}{4a\sqrt{P(a)}}
\right)
\end{equation}
where $x_{i}$ and $a_{i}$ stands for the initial values at time $t_{i}$.

\subsection{Qualitative analysis}\label{Analysis}
The shell expansion is analysed in term of dimensionless quantities~: the  {\it magnification\/} $X$ and the {\it expansion rate\/} $Y$
\begin{equation}\label{XY}
X=\frac{x}{x_{i}},\qquad Y=\frac{\alpha}{H_{\circ}}
\end{equation}
Their evolution with the expansion parameter $a$ are obtained from eq.\,(\ref{impulsionC2}, \ref{Croissance}) by numerical integration\footnote{Because the mapping $t\mapsto a(t)=1/(1+z)$ is a monotonic function  in the present investigation, their evolution with cosmic time $t$ or with redshift $z$ can be straightforwardly derived.}. Let us focus our investigation around the generally accepted values $\lambda_{\circ}=0.7$ and $\Omega_{\circ}=0.3$. The initial conditions lie on the expansion rate $Y_{i}$ and the formation date $t_{i}$, as expressed by means of $a_{i}=a(t_{i})$. The values $a_{\rm i}=0.003$ and $Y_{\rm i}=0$ (void initially expanding with Hubble flow) are used as standard in our discussion. Hereafter, we simply provides us with synthesis results, more information can be found in \cite{FT06}.

\subsubsection{Kinematics}\label{kinematics}
\begin{figure}
\includegraphics[width=3.5in]{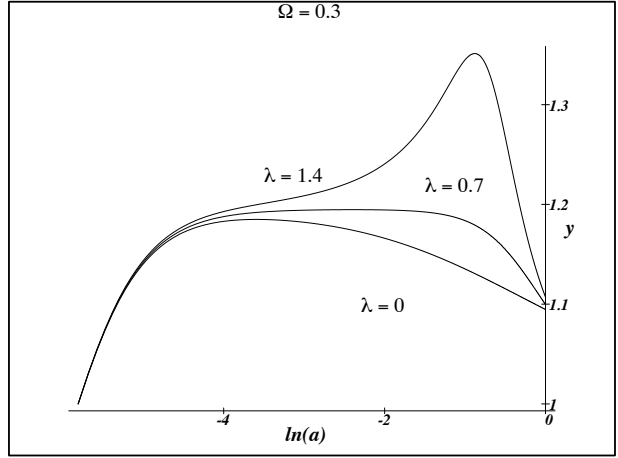}
\caption{The corrective factor $y$ to Hubble expansion.}
\label{Fig0}
\end{figure}

The expansion velocity of the shell S with respect to its centre reads 
\begin{equation}\label{vitesse}
\vec{v}=y H \vec{r},\qquad y=1+\frac{Y}{h},\qquad h=\frac{H}{H_{\circ}}
\end{equation}
where $y$ stands for the corrective factor  to Hubble expansion. The general trend of the kinematics can be derived from the diagram $y$ versus $a$ in Fig.\,\ref{Fig0}. It results from the initial conditions $a_{\rm i}=0.003$,  $Y_{\rm i}=0$, and the cosmological parameters $\Omega_{\circ}=0.3.$ and  $\lambda_{\circ}=0$,  $\lambda_{\circ}= 0.7$, $\lambda_{\circ}=1.4$.

The shell S expands faster than Hubble expansion ($y>1$) at early stages of its evolution with no significant dependence on $\Lambda$. $\Lambda$-effect appears later by preserving the curve from an earlier decreasing. If $\lambda_{\circ}>0.7$ ({\it i.e.\/}, $k_{\circ}>0$) then it is characterized by a significant protuberance at $z\sim 1.7$, the larger the $\lambda_{\circ}$ ({\it i.e.\/}, $k_{\circ}$) the higher the bump. It is due to the existence of a minimum value  $H_{m}$ of Hubble parameter $H$ which is reached during the cosmological expansion, see eq.\,(\ref{extremum}). After this period the expansion of S reaches asymptotically Hubble behavior. It is interesting to note that the present epoch ($a=1$) appears quite peculiar because of the relative proximity of curves, but it is solely an artifact\footnote{Indeed, the three curves cross at $a>1$ but not all in only one point.}.

\subsubsection{Dichotomy between cosmological parameters}\label{dependence}
With the aim of disentangling the effect of the cosmological constant on the evolution of spherical voids from that of the outer density (herein named $\Lambda$-effect and $\Omega$-effect respectively), the dependence of $X$ and $Y$ on cosmological parameters $\Omega_{\circ}$ and  $\lambda_{\circ}$ are investigated separately within acceptable values domains in accordance with observations.
 
\begin{enumerate}
\item $\Omega$-effect.\label{OmegaVoid} ---
With a constant $\lambda_{\circ}=0.7$, the magnification $X$ and the expansion rate $Y$ increase with $\Omega_{\circ}$, what interprets by the attraction of shell particles toward denser regions. The growth shows a huge burst which freezes asymptotically up to matching Hubble expansion ($Y=0$). This trend is not significantly modified by other acceptable values of $\lambda_{\circ}$.

 \item $\Lambda$-effect .\label{LambdaVoid} ---
With a constant $\Omega_{\circ}=0.3$,  the larger the $\lambda_{\circ}$ the larger the magnification X. As for $\Omega$-effect, the magnification $X$ increases nonlinearly with $\lambda_{\circ}$. With $\Lambda>0$, this phenomenon interprets as a repulsive effect of gravity in empty regions (also named gravitational repulsion of vacuum). The expansion rate $Y$ is weakly dependent on  $\lambda_{\circ}$, what means that it does not characterizes $\Lambda$. However, it has a cumulative late effect which is reflected on $X$.

\item Interpretation of parameter $k_{\circ}$.\label{kVoid} ---
According to eq.\,(\ref{void}), the test-particle within S moves in the reference frame toward its centre but if $k_{\circ}>0$ then it starts moving toward its border at date $a(t^{\star})=\frac{8}{3} \pi{\rm G}\rho_{\circ}K_{\circ}^{-1}$. Such a property of sweeping out the void region (due to $H<H_{\infty}$) interprets as a stability criterion for void regions. A dimensional analysis of eq.\,(\ref{K}) shows that the Newtonian interpretation of $k_{\circ}=\lambda_{\circ}+\Omega_{\circ}-1$ corresponds to a dimensionless binding energy for the universe\footnote{The lower $k_{\circ}$ the faster the cosmological expansion. Note that it works with an opposite direction for the expansion of spherical voids.}. Because $K_{\circ}$ is an integration constant, this provides us with a meaningful procedure for comparing the magnitudes of $\Lambda$-effect or $\Omega$-effect between different world models at constant $k_{\circ}$. For Friedmann model,  $k_{\circ}$ stands for the dimensionless spatial curvature (of the comoving space), $k_{\circ}>0$ world models correspond to spatially closed universes.

\item  Dependence on initial conditions. \label{InitCind} ---
The earlier the date of birth the larger the magnification $X$. One has a limiting curve $a\mapsto Y$ defined by $a_{i}=0$, which characterizes $\Omega_{\circ}$. All expansion rate curves related to other formation dates are located in lower part and reach it asymptotically. 
One can expect that the effects on the dynamics of growth resulting from reasonable initial expansion rates $Y_{i}\neq 0$, related to physical processes ({\it e.g.\/}, supernovae explosions), are negligible at primordial epoch\footnote{Since Hubble expansion is all the more important towards the past, the earlier the formation date the weaker this effect, according to eq.\,(\ref{vitesse}).}, what legitimizes the initial condition $Y_{i}=0$ at $a_{i}=0.003$. Evolutions with other initial conditions on $Y_{i}$ can be deduced from that since any given point $(a_{i},Y_{i})$ in the diagram belongs to a single evolution curve.
\end{enumerate}

\subsubsection{Synthesis}\label{Synthesis}
As a result, the expansion of a spherical void does not show a  {\it linear regime\/}, but a huge initial burst which freezes asymptotically up to matching Hubble expansion. Its radius  (in the reference frame) $x$ increases from an initial size $x_{i}$ with the cosmological parameters $\Omega_{\circ}$ and $\lambda_{\circ}$. The related individual effects interpret respectively by the gravitational attraction from the outer parts and by the gravitational repulsion of its borders (or the vacuum from the inner parts). The larger these parameters  the higher the magnification $X=x/x_{i}$. The dynamics is sensitive to $\Omega$-effect at primordial epochs and to $\Lambda$-effect later on. The evolution of its expansion velocity (in the reference frame) $\vec{v}_{c}=YH_{\circ}\vec{x}$ with time does not characterise $\lambda_{\circ}$ but $\Omega_{\circ}$. On the other hand, the cosmological constant intervenes significantly on the kinematics by means of the corrective factor $y$ to Hubble expansion by preserving the expansion from an earlier decreasing. Moreover, a values domain of cosmological parameters, which corresponds to relativistic
 spatially closed world models, are characterised by a significant expansion of voids. The later reaches a maximum at redshift $z\sim 1.7$ (with $\Omega_{\circ}\sim 0.3$), the larger the $\lambda_{\circ}$ ({\it i.e.\/}, $k_{\circ}$) the higher the expansion rate. Notwithstanding, the perturbation on redshift of sources located on the shell of expanding voids does not exceed 
\begin{equation}\label{redshift}
\Delta\,z= \frac{XY}{1+z} \frac{x_{i}H_{\circ}}{c}
\end{equation}
which is a tiny value ($\sim 10^{-3}$) because of counterbalancing behaviors of $X$ and $Y$. The interior of voids shows a de Sitter expansion $\vec{v}=\sqrt{\Lambda/3}\vec{r}$ which sweeps out it  if $k_{\circ}>0$, what  interprets as a stability criterion. The expansion rate evolutions related to formation dates show a common envelope curve which characterizes the density parameter $\Omega_{\circ}$.

\subsection{Discussion}\label{Discussion}
It is clear that such a simple model should be applied with caution to observed distribution of galaxies since the effects related to foamlike patterns are not considered. Moreover, because Newtonian approach is used instead of general relativity, it applies solely to reasonable sized voids\footnote{At first glance,  on might argue that the related scale should be larger than the size of background structures in order to make the uniform density hypothesis an acceptable approximation, what looks like a dilemma. However, such an hypothesis, which applies to the entire universe, is ensured by the isotropy of CMB.} so that the main features should remain qualitatively reliable. On the other hand, based on the connection of two exact solutions of Euler-Poisson equations system, which is a huge advantage for investigating properly the behavior of a single void, it provides us with an appreciable hint to the dynamics with respect to cosmological parameters.

\section{Conclusion}\label{Conclusion}

Euler-Poisson equations system written in an adapted coordinates system enables us to define in a straightforward way cosmological solutions, such as Newton-Friedmann and the vacuum (de Sitter) models, to investigate an an-isotropic extension of Hubble flow and the dynamics of spherical voids in an expanding universe, as long as Newton approximation with cosmological constant is valid. The related results are summarised as follows~:
\begin{itemize}
\item {\it An-isotropic Hubble flow\/}. The present solution generalizes the Hubble law and provides us with a better understanding of cosmic velocity fields within large scale structures. As a result, it implies necessarily an homogenous distribution of gravitational sources, as similarly to Hubble law. Because the chronology identifies to FL chronology for a vanishing distortion, this model interprets as a Newton approximation of anisotropic cosmological solutions. The motions are characterised by means of a constant of motion $\alpha$. Among them, particular solutions can be easily derived for $\alpha=0$. They describe all planar distortions, in addition of two classes of planar-axial distortions with or without rotation. Among these solutions, the one which ensures a planar kinematics is of particular interest because it describes constant (eternal) and rotational distortions. This solution can be fully determined from observational data except for the orientation of the rotation. The sensible result is that the velocity field is not potential. It is interesting to note that this model accounts for motions which might be interpreted as due to tidal forces whereas the density is homogeneous. It is an alternative to models which assume the presence of gravitational structures similar to Great Attractor as origin of a bulk flow.

\item {\it Spherical voids\/}. The model of an isotropic universe constituted by a spherical void surrounded by a uniform distribution of dust is defined such that the behaviors of bordering regions inside and outside the shell, coincide with Newton-Friedmann (NF) solutions. The connection conditions between these two regions are investigated in Classical Mechanics, what provides us with the dynamics of the shell. Such a schema does not correspond to the usual embedding of a void solution into a cosmological background solution, but interprets as a non linear perturbation of NF models. The general behavior of the void expansion shows a huge initial burst, which freezes asymptotically up to match Hubble expansion. While the corrective factor to Hubble law on the shell depends weakly on cosmological constant at early stages, it enables us to disentangle significantly cosmological models around redshift $z\sim 1.7$. The magnification of spherical voids increases with the density parameter and with the cosmological constant. An interesting feature is that for spatially closed Friedmann models, the empty regions are swept out, what provides us with a stability criterion.
\end{itemize}

\appendix
\section{Appendix : Covariant formulation of Euler equations system}\label{Covariant}
Souriau's covariant formulation of Euler-Poisson equations\cite{Souriau70} can be summarized as follows~: The geometrical interpretation of Newton dynamics\cite{DuvalKunzle78} shows that the component of gravitational field $\vec{g}$ identify to the only non null Christoffel components of newtonian connection 
$$
\Gamma^{j}_{00}=-g^{j}
$$
Hence, one obtains their expression in the new coordinates system defined in eq.\,(\ref{tx}),  and the only non null components read
$$
(\Gamma_{c})^{j}_{k0}=H \delta^{j}_{k},\qquad (\Gamma_{c})^{j}_{00}=-g_{c}^{j}
$$
Let ${\cal T}$ be a function(al) defined on the set of symmetric density tensors $x\mapsto \gamma$ on newtonian spacetime $R^{4}$
$$
{\cal T}(x\mapsto\gamma)=\int_{R^{4}}T^{\mu\nu} \gamma_{\mu\nu}{\rm d}t{\rm d}V
$$
where ${\rm d}t{\rm d}V$ stands for the volume element and $T$ for a symmetric contravariant tensors  which accounts for the media. The measure density ${\cal T}$ is eulerian if and only if it vanishes for all covariant tensor fields  which reads
$$
\gamma_{\mu\nu}=\frac{1}{2}\left( \hat{\partial}_{\mu} \xi_{\nu}+ \hat{\partial}_{\nu} \xi_{\mu}\right)
$$
where $\hat{\partial}$ stands for the covariant derivative and $x\mapsto\xi$ for a \underline{compact support} 1-form.
In such a case, it is obvious to show that 
$$
\hat{\partial}_{\mu}T^{\mu\nu}=0
$$
interprets as Euler equations, in any coordinates systems.

\begin{theacknowledgments}
The authors thanks warmly M\'ario Novello, and co-organizers, for this high level  School, the  hospitality and financial supports.
\end{theacknowledgments}

\end{document}